\documentclass[aps,twocolumn,pra,tightenlines,floatfix]{revtex4}
\usepackage[dvips]{graphicx}
\usepackage[english]{babel}
\usepackage{amsmath}
\usepackage{amssymb}
\usepackage{times}

\newcommand{\bs}{\begin{split}}
\newcommand{\es}{\end{split}}
\newcommand{\be}{\begin{equation}}
\newcommand{\ee}{\end{equation}}
\newcommand{\ba}{\begin{eqnarray}}
\newcommand{\ea}{\end{eqnarray}}
\newcommand{\mb}[1]{{\mathbf{#1}}}

\newcommand{\ek}{\epsilon_{\mathbf{k}}}
\newcommand{\Ek}{E_{\mathbf{k}}}

\begin{document}

\title{Temperature and final state effects in radio frequency
  spectroscopy experiments on atomic Fermi gases}

\author{Yan He, Chih-Chun Chien, Qijin Chen, and K. Levin}

\affiliation{James Franck Institute and Department of Physics,
 University of Chicago, Chicago, Illinois 60637}

\date{\today}

\begin{abstract}
  We present a systematic characterization of the radio frequency (RF)
  spectra of homogeneous, paired atomic Fermi gases at finite
  temperatures, $T$, in the presence of final state interactions.  The
  spectra, consisting of possible bound states and positive as well as
  negative detuning ($\nu$) continua, satisfy exactly the zeroth- and
  first-moment sum rules at all $T$.  We show how to detect the $\nu <
  0$ continuum arising from thermally excited quasiparticles, which has
  not yet been seen experimentally.  We explain semi-quantitatively
  recent RF experiments on ``bound-bound'' transitions and, thereby,
  predict the associated effects of varying temperature.
\end{abstract} 
\pacs{03.75.Hh, 03.75.Ss, 74.20.-z \vspace*{-2ex}
}

\maketitle

The superfluid and normal phases in trapped Fermi gases undergoing BCS
to BEC crossover are presenting us with novel forms of superfluidity.
An important characteristic of the superfluid is the pairing gap which
is best probed using radio frequency (RF) spectroscopy
\cite{Grimm4,Ketterle4}.  This technique has been applied
experimentally in a trap integrated \cite{Grimm4,KetterleRF} and
tomographic \cite{Ketterle4} fashion.  While early theoretical work
\cite{Torma2,heyanall} addressed trap effects, more recently attention
has been on final state effects \cite{Baym2,Punk,Strinati7,Basu,Torma3}
although, unfortunately, only at low or zero temperature.  Many have
viewed the importance of these experiments as a means of quantitatively
measuring the ground state pairing gap, thereby testing different
approaches to BCS-BEC crossover. Our point of view is that finite
temperature is crucial to full experimental understanding as well as
reliable assessments of theory.

The goal of this paper is to present a single formalism for the RF
spectra at all frequencies and all $T$, including final state effects.
A successful theory of a Fermi gas near unitarity, not only (i) has a
pairing gap which appears \cite{heyanall,Torma2,Punk} at $T^* > T_c$ but
which, as $T$ is decreased, (ii) exhibits a second order phase
transition, at $T_c$.  Studies of this smoothly varying (from above
$T^*$ to $T=0$) pairing gap, reminiscent of its counterpart in the high
$T_c$ superconductors, may elucidate some of the physics of the cuprates
\cite{ourreview}.  On physical grounds \cite{heyanall,Torma2,Punk} it is
clear that the RF current $I(\nu)$ reflects the pairing gap $\Delta(T)$
rather than coherent superfluid order.  At odds with this observation is
the fact that all crossover theories which include pairing fluctuations
\cite{Zwerger,Strinati8,Griffingroup2,Drummond3} except the present one,
lead to first order transitions at $T_c$.  In a related fashion,
alternative calculations \cite{Strinati7,Stoof3,Basu} of $I(\nu)$
consider only the low or zero $T$ superfluid and/or separately the
normal phase even though, at $T < T^*$, the presence or absence of
superfluid order in the RF spectra should not lead to fundamentally
different physics.

We consider a homogeneous system which is relevant to recent tomographic
experiments \cite{Ketterle4} At $T \neq 0$, the spectrum consists of
(possibly) bound state contributions which either appear at positive or
negative detuning, $\nu$ and, (always), positive as well as negative
$\nu$ continuum contributions which reflect the pairing gap, and can be
used to measure its size.  We emphasize the $\nu <0$ continuum which
derives from thermally excited quasiparticles has not yet been seen
experimentally nor addressed theoretically. A central finding is that it
can be strongly enhanced by final state interactions and made visible in
future tomographic experiments.  Near unitarity, final state effects
make it possible to extract (using sum rules) the gap $\Delta$ as well
as the chemical potential $\mu$. We explain semi-quantitatively recent
low $T$ experiments and make predictions for the accompanying
temperature dependences which should be observable.

The RF technique focuses on the three lowest energy atomic hyperfine
states (two of which are involved in the pairing, while a third provides
a final ``excited'' state for one component of a pair).  For
definiteness, we first consider a superfluid of pairs in the equally
populated hyperfine 1-2 levels and apply a radio frequency $\omega_{23}$
to excite the atoms in state 2 to state 3, as described by a Hamiltonian
given in Refs.~\cite{heyanall,Torma2,Baym2,Strinati7,Basu,Punk}. The RF
response function can be obtained following the standard linear response
theory \cite{heyanall,Torma2,Baym2}.  Here we formulate the finite $T$,
RF problem using a diagrammatic scheme which can be made compatible with
the diagrams in Ref.~\onlinecite{Strinati7}, although attention in that
paper was restricted to very low temperatures.  We will see below that
our diagrammatic scheme reduces at $T=0$ to the approach of
Ref.~\onlinecite{Basu}.  This correspondence, and indeed, all
diagrammatic formulations \cite{Punk} of the RF experiments are based on
a $T$-matrix approach.  The $T$-matrix used here (for the 1-2 channel)
is consistent \cite{Chen2,ourreview} with the BCS-Leggett ground state
equations and involves one bare and one dressed Green's function.  We
have
\begin{eqnarray}
t^{-1}_{12}(Q)&=&g_{12}^{-1}+\sum_K G_1(K)G_2^0(Q-K)\\
t^{-1}_{13}(Q)&=&g_{13}^{-1}+\sum_K G_1(K)G_3^0(Q-K)
\end{eqnarray}
where we have introduced the dressed Green's function $G = [ (G^0)^{-1}
- \Sigma ]^{-1}$ and $G^0$ is the Green's function of the
non-interacting system. Here the subscripts indicate the hyperfine
levels, $K\equiv (i\omega_l, \mathbf{k})$, $Q\equiv (i\Omega_n,
\mathbf{q})$ are 4-momenta with $\sum_K \equiv T\sum_{l}
\sum_{\mathbf{k}}$, etc., and $\omega_l$ and $\Omega_n$ are fermion and
boson Matsubara frequencies, respectively. Throughout we take $\hbar =
k_B =1 $ and assume a contact potential (so that the strict Hartree
self-energy vanishes) and a (nearly) empty population in the hyperfine 3
state so that $G_3(K)\approx G_3^0(K)$.  As has been demonstrated
elsewhere \cite{ourreview}, it is reasonable to take the self energy (on
the real frequency axis) in the Green's functions $G_1$ and $G_2$ to be
of the generalized BCS form
\begin{equation}
\Sigma(\omega,\mb{k}) \approx \frac{\Delta^2}{\omega +\ek} \,,
\label{Sigma}
\end{equation}
although this approximation is not essential.  Similarly, we have shown
\cite{ourreview} that, below $T_c$, $\Delta(T)$ is constrained by a
BCS-like gap equation which can be written as $1+ g_{12} \chi_{12}(0) =
0$ where $\chi_{12}(Q) = \sum_K G_1(K)G_2^0(Q-K)$, in conjunction with a
fermion number equation.  More generally, the propagator for
noncondensed pairs is of the form
$t_{12}(Q) = g_{12}/[1 + g_{12} \chi_{12}(Q)]$.

We emphasize a distinction between the pairing gap (which we call
$\Delta$) and the order parameter, called $\Delta_{sc}$.  The difference
between these two energy scales can be shown \cite{ourreview} to be
associated with noncondensed pair effects parameterized by the pseudogap
$\Delta_{pg}$ defined by
\begin{equation}
\Delta_{pg}^2(T)  = \Delta^2 (T) - \Delta_{sc}^2 (T)\,.
\label{eq:Dpg}
\end{equation}
Here we note that $\Delta_{pg}^2 = -\sum_{Q\neq 0} t_{12}(Q)$, which
allows $T_c$ to be determined \cite{ourreview} as the temperature where
$\Delta_{sc}$ first vanishes. We find $T_c = 0.25 T_F$ at unitarity.
It is convenient notationally to define a form of Gor'kov $F$ function
in terms of the pairing gap as
$$
\Delta G_2(K)G^0_1(-K)
=\frac{\Delta}{\omega_l^2 +E_k^2} \equiv F(K)\,,
$$
where $E_k = \sqrt{ \xi_k^2 + \Delta^2 (T) }$, and $\xi_k = \epsilon_k
-\mu$, $\epsilon_k=k^2/2m$.
Because of the constraints imposed by the BCS-like gap equation, $t_{12}(Q)$
diverges at $Q=0$ so that it is reasonable to set $Q$ in $t_{12}$ to
zero, i.e., $t_{12}(Q) \approx -(\Delta^2/T)\delta(Q)$. This assumption, which
leads to the simple form of Eq.~(\ref{Sigma}), is not essential for
understanding the physics but it does greatly simplify the calculations
\cite{mean-field-footnote}.

The resulting diagram set for the RF response function, $D(Q)$, is shown
in Fig.~1 and this last approximation is equivalent to treating the
Aslamazov-Larkin (AL) diagram (called $D_{AL}$) in Fig.~1 at the BCS
mean-field level, leading to the opposite momenta $\pm K$ for particles
1 and 2 in the diagram.  The leading order term, $D_0(Q)$, of the
response function appears as the bubble on the left and was introduced
in Ref.~\onlinecite{Torma2}. The term on the right, $D_{AL}(Q)$, depends
on $\Delta$, not $\Delta_{sc}$, and incorporates final-state effects via
the interactions $g_{12}$ between 1 and 2 and $g_{13}$ between 1 and 3.
We neglect the effects arising from the interaction between 2 and 3.
This is consistent with the approach in Ref.~\onlinecite{Baym2}. This
second term has appeared previously in studies of the superfluid density
\cite{Chen2}.

\begin{figure}
\includegraphics[width=3.3in,clip]{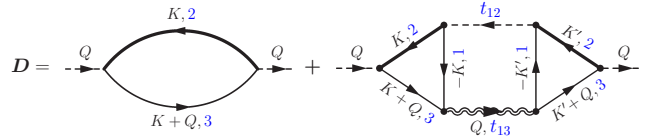}
\caption{(Color online) Feynman diagrams for the RF response function
  $D(Q)$. The left bubble is the lowest order $D_0$, whereas the right
  diagram, $D_{AL}$, is associated with final state effects. Here thin
  (thick) lines stand for bare (full) fermion propagators, the dashed
  line for $t_{12}$, approximated as the condensate, and double wiggly
  line for $t_{13}$. The numbers in blue indicate the hyperfine levels.}
\label{fig:1}
\end{figure}

Writing out the AL diagram yields
\begin{eqnarray}
D_{AL}(Q)=\Big[\sum_K F(K)G_3^0(K+Q)\Big]^2 t_{13}(Q)\,.
\label{eq:D_AL}
\end{eqnarray}
For the RF field, $Q=(i\Omega_n, \mathbf{0})$ so that
$D(i\Omega_n)\equiv D(Q)$. We take $\mu_{3}$ satisfying
$f(\xi_{k,3})=0$, where $\xi_{k,3}=\epsilon_k-\mu_3$. Then the RF
current, given by the retarded response function, is $I(\nu) \equiv
-(1/\pi)\,\mbox{Im}\,D^R(\Omega)$, where $\Omega\equiv\nu+\mu-\mu_3$,
and we find
\begin{equation}
  D(Q )  =D_0 (Q) +  \frac{[D_2(Q)]^2} {m/4\pi a_{13} + D_1(Q)}\,, 
\label{eq:D}
\end{equation}
and $t_{13}^{-1} (Q) = m/4\pi a_{13} + D_1(Q)$, where $a_{13}$ (and
$a_{12}$) are the $s$-wave scattering length in the 1-3 (and 1-2)
channels, respectively.
Here $D_0(Q) = \sum_K G_2(K)G_3^0(K+Q)$
\begin{equation}
 = \sum_K\Big[\frac{f(E_k)-f(\xi_{k,3})}{i\Omega_n\!+E_k-\xi_{k,3}}u_k^2 
  +\frac{1\!-\!f(\xi_{k,3})\!-\!f(E_k)}{i\Omega_n-E_k-\xi_{k,3}}v_k^2\Big]
\end{equation}
and $I_0(\nu)=-(1/\pi)\,\mbox{Im}\,D_0^R(\Omega)$. We also define 
$D_2(Q) \equiv \sum_K F(K)G_3^0(K+Q)$ 
\begin{equation}
  =\sum_K\frac{\Delta}{2E_k}\Big[\frac{1\!-\!f(E_k)\!-\!f(\xi_{k,3})}
  {i\Omega_n-E_k-\xi_{k,3}} 
  -\frac{f(E_k)-f(\xi_{k,3})}{i\Omega_n\!+E_k-\xi_{k,3}}\Big]
\end{equation}
and
$D_1(Q) \equiv \sum_K G_1(K)G_3^0(Q-K)-\sum_\mathbf{k} (1/2\epsilon_k) =$
\begin{equation}
\sum_K\Big[\frac{f(E_k)\!+\!f(\xi_{k,3})\!-\!1}{i\Omega_n-E_k-\xi_{k,3}}u_k^2
+\frac{f(\xi_{k,3})-f(E_k)}{i\Omega_n\!+\!E_k\!-\!\xi_{k,3}}v_k^2\Big]
-\sum_\mathbf{k} \frac{m}{k^2}\,.
\label{eq:D1}
\end{equation}
After analytical continuation and change of variables, we have $\Omega
\pm \Ek -\xi_{k,3} = \nu \pm \Ek -\xi_k$. Importantly, the
denominators here are the same as those which appear in
$t_{12}$. Furthermore, at $\nu=0$, $f(\xi_{k,3})$ is cancelled out so that
\begin{equation}
  t^{-1}_{13}(0) =
  (g_{13}^{-1} - g_{12}^{-1}) + t_{12}^{-1}(0) = g_{13}^{-1} - g_{12}^{-1}.
\end{equation}
It follows that the complex functions $D_0(Q)$, $D_1(Q)$, and $D_2(Q)$
are the same as their wave function calculation counterparts \cite{Basu}
when the pairing gap $\Delta$ is chosen to be order parameter
$\Delta_{sc}$ and $T=0$. It is $\nu$ not $\Omega$ that should be
identified with the experimental RF detuning.

After some straightforward algebra, one can show that when $g_{13} =
g_{12}$ there is an exact cancellation such that $I(\nu)\sim \delta(\nu)$.
In general, we have $I_0(\nu) = (1/\pi)(\Delta^2/\nu^2)\, \mbox{Im}\,
\bar{t}^{-1,R}_{13}(\nu) $, and
\begin{eqnarray}
\lefteqn{I(\nu) = \left[\frac{1}{g_{12}} - \frac{1}{g_{13}}\right]^2\!\!
\frac{I_0(\nu)}{|\bar{t}^{-1,R}_{13}(\nu)|^2}}\nonumber\\
&&= -\frac{1}{\pi} \left[\frac{m}{4\pi a_{13}} - \frac{m}{4\pi a_{12}}\right]^2
\!\frac{\Delta^2}{\nu^2}\, \mbox{Im}\, \bar{t}^R_{13}(\nu),
\label{eq:I}
\end{eqnarray}
where $\bar{t}^R_{13}(\nu)\equiv t^R_{13}(\Omega)$.

Equations (\ref{eq:I}) are a central result of this paper which make it
clear that final state effects in the RF current directly reflect the
$T$-matrix in the 1-3 channel.  In general, features in the RF spectra
derive from the poles and imaginary parts of
Eqs.~(\ref{eq:D})-(\ref{eq:D1}).  The spectrum may contain a bound state
associated with poles at $\nu_0$ in $t_{13}$, as determined by $
t^{-1}_{13}(\nu_0) = 0$.  This leads to the so called ``bound-bound''
transition. In addition, there is a continuum associated with both the
numerator and denominator in the first of Eqs.~(\ref{eq:I}), with each
contribution spanned by the limits of $\nu = \xi_k\pm E_k$, i.e.,
$-(\sqrt{\mu^2 + \Delta^2} +\mu) \le \nu \le 0$ and $\nu \ge \sqrt{\mu^2
  + \Delta^2} -\mu $.  The continuum at positive frequencies is
primarily associated with breaking a pair and promoting the state 2 to
state 3. This represents the so-called ``bound-free'' transition. On the
negative detuning side, the continuum is primarily associated with
promoting to state 3 an already existing thermally excited 2
particle. The spectral weight of the negative continuum vanishes
exponentially at low $T$ as $e^{-\Delta/T}$.  Therefore, there is a
strong asymmetry in the continuum with the bulk of the weight on the
positive frequency side for low $T$.  If the bound state falls within
the negative continuum, it will acquire a finite life time, and decay
quickly at high $T$.

\begin{figure}
\includegraphics[width=3.2in,clip]{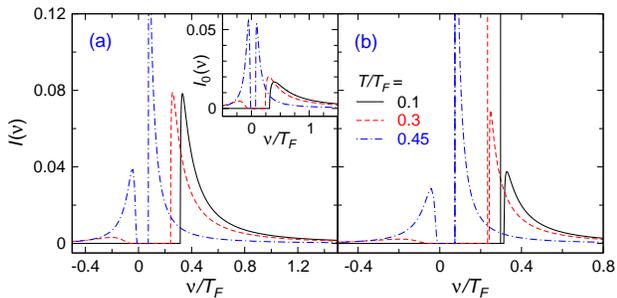}
\caption{(Color online) RF current $I(\nu)$ as a function of RF detuning
  $\nu$ for transitions from unitarity $1/k_Fa_{12}=0$ at 834~G to final
  state (a) $1/k_Fa_{13}=-1$ and (b) $-0.5$ in the BCS regime,
  corresponding to $T_F = 31$ and 124~kHz, respectively. The
  temperatures are $T/T_F=0.1$ (Black solid), $0.3$ (red dashed) and
  $0.45$ (blue dot-dashed lines). Here $T_c=0.25T_F$.  The sharp lines
  next to the right continuum in (b) correspond to bound states.  Inset:
  Lowest order RF current $I_0(\nu)$ vs $\nu$.  }
\label{fig:2}
\end{figure}

Of importance, in assessing a theoretical framework for computing the RF
current are the two sum rules associated with the total integrated
current and the first moment or ``clock shift'' \cite{Baym2}. Using the
Kramers-Kronig relations between $\mbox{Re}\,t^R_{13}$ and
$\mbox{Im}\,t^R_{13}$, it is easy to prove that, not only in the ground
state, but also at finite temperature, Eq.~(\ref{eq:I}) satisfies
\begin{eqnarray}
  \int d\nu  \, I(\nu) &=& n_2 - n_3 \,,\\
  \int\! d\nu \,\nu\, I(\nu) 
&=& \Delta^2 
\frac{m}{4\pi}\, \left(\frac{1}{a_{12}} - \frac{1}{a_{13}}\right)\,,
\end{eqnarray}
where $n_2$ and $n_3 (=0)$ are the density of state 2 and 3 atoms,
respectively.  In this way we find for the clock shift
\begin{equation}
  \bar{\nu} = \frac{\int d\nu \,\nu I(\nu)} {\int
    d\nu \,I(\nu)}
  =\frac{ \Delta^2}{n_2 - n_3}
  \frac{m}{4\pi} \left(\frac{1}{a_{12}} - \frac{1}{a_{13}}\right)\,,
\label{eq:14}
\end{equation}
which agrees with Ref.~\cite{Baym2} when $n_3 \rightarrow 0$.
It should be stressed that this sum rule is satisfied only when
$a_{13}\neq 0$ and when both diagrammatic contributions are included. It is
easy to show that at large $\nu$, $I_0(\nu) \sim \nu^{-3/2}$,
$\mbox{Im}\,t_{13}^R \sim \nu^{-1/2}$, so that $I(\nu) \sim \nu^{-5/2}$,
in agreement with Ref.~\onlinecite{Strinati7}.  Clearly, the first
moment of $I(\nu)$ is integrable, whereas the first moment of $I_0(\nu)$
is not.  Finally, Eq.~(\ref{eq:I}) reveals that the spectral weight
(including possible bound states) away from $\nu =0$ will disappear when
the gap $\Delta$ vanishes.

Figures 2(a) and 2(b) illustrate the behavior of the spectrum $I(\nu)$
when the initial state 1-2 pairing is at unitarity (i.e., at 834~G) and
the final state 1-3 pairing is on the BCS side of the 1-3 resonance, for
temperatures $T/T_F =0.1$, 0.3, and 0.45.  The parameters we use are
taken from Ref.~\onlinecite{ChinJulienne}. The inset of Fig.~2(a)
indicates the behavior in the absence of final state effects for the
same temperatures. The asymmetry of the continuum around $\nu =0$,
discussed earlier, is evident even in this leading order bubble diagram.
As $T$ is raised the spectrum becomes more symmetric.  In contrast to
the findings in Ref.~\onlinecite{Stoof3}, and as consistent with
experiments \cite{Grimm4} on trapped gases, we do not find a substantial
pairing gap at $T/T_F \approx 1$.  In Fig.~2(a), the final state
interaction $1/k_Fa_{13}=-1$ is relatively weak, and there is no bound
state. In contrast, at $1/k_Fa_{13}=-0.5$ (or $T_F \approx 6\mu$K) in
Fig.~2(b), a bound state emerges at low $T$ (although it disappears at
moderate temperatures when the gap becomes small). For the low $T_F\sim
2.5\mu$K used in Ref.~\onlinecite{Grimm4}, we do not find a bound
state. These results are consistent with $ T = 0$ calculations of Basu
and Mueller \cite{Basu}.  It should be stressed that, at 834~G for a
typical $T_F$, when the bound-bound transition occurs, it is barely
separated from the asymmetric bound-free continuum, which is always
present.

\begin{figure}
\includegraphics[width=3.3in,clip]{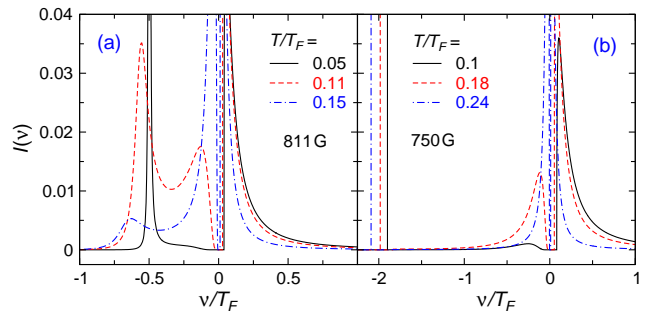}
\caption{(Color online) RF current $I(\nu)$ as a function of detuning
  $\nu$ for $1\rightarrow 2$ transitions in a 1-3 superfluid of $T_F =
  40$~kHz (a) from $1/k_Fa_{13}=-0.804$ to final states $1/k_Fa_{23}=0$
  at 811~G, and (b) from $1/k_Fa_{13}=-0.524$ to $1/k_Fa_{23}=0.68$ at
  750~G, for different temperatures as labeled.  Here $T_c/T_F = 0.15$
  and 0.17, respectively. In  (b) when $T$ is high and $\Delta$ is
  small, the two peaks around $\nu = 0$ may not be resolvable
  experimentally.}

\label{fig:3}
\end{figure}

Figure 3 presents the analogous plots at different $T$ for RF
transitions from an initial 1-3 superfluid with $T_F = 40$~kHz at (a)
811 and (b) 750~G, which are on the BCS side of the 1-3 resonance (which
appears at 690~G). The system is subject to an RF field promoting state
1 to state 2.  ``Bound-bound-like transitions'' \cite{ChinJulienne} now
appear. In Fig.~3(a), the bound state falls within the negative detuning
continuum.  Importantly, the disappearance of the bound state with
temperature is preceded by a very unusual two-peaked spectrum in the
negative detuning regime, which is seen at the two higher $T$.  We can
understand this unusual structure as a combination of the peak from the
negative continuum which appears very close to $\nu =0$, (as also seen
in Figure 2) and the near-by bound state peak.  At even higher $T$, the
spectral weight will shift almost completely to the region near $\nu
=0^-$ and the bound state decays rapidly.  As $\nu \rightarrow 0^-$, the
negative continuum peak is a combined effect of the vanishing $\mbox{Im}
\bar{t}^R_{13}$ and the diverging factor $1/\nu^2$ in Eq.~(\ref{eq:I}).
In Fig.~3(b), the bound state is outside the continuum, and the binding
energy is fairly insensitive to temperature.  We have chosen
experimentally accessible parametesrs here, so that the unusual
double-peaked structure in $I(\nu)$ at $\nu<0$ should be observable.
Finally, we emphasize that the highest $T$ cases in Figs. 2 and 3 are at
or above $T_c$, so that the continuum appears only because there exists
a pseudogap in the fermionic spectrum.

\begin{figure}
\includegraphics[width=3.3in,clip]{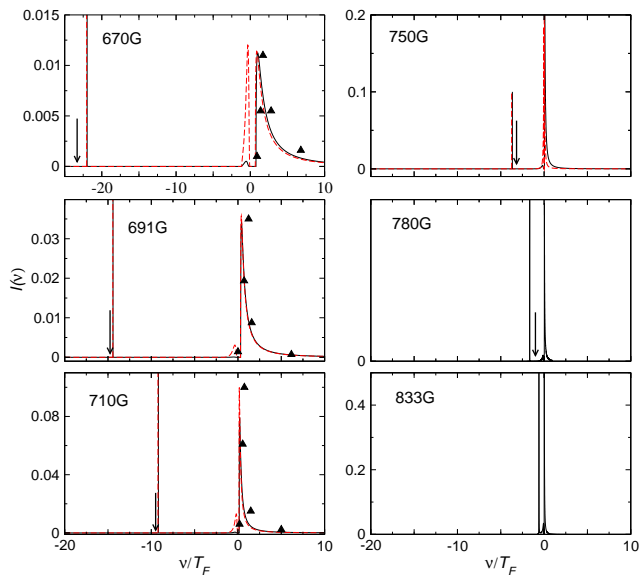}
\caption{(Color online) RF current $I(\nu)$ as a function of detuning
$\nu$ for a 1-3 superfluid with RF excitation from state 3 to state 2.
The black curves are calculated at experimental parameters of
$(1/k_Fa_{13}, 1/k_Fa_{12}, T/T_F) =(0.4, 3.3, 0.2)$, $(0.0,
2.6, 0.1)$, $(-0.3, 2.0, 0.1)$, $(-0.7, 1.1, 0.09)$,
$(-0.9, 0.6, 0.09)$, and $(-1.2, 0.0, 0.06)$
from low to high fields.  The red dashed curves are calculated at twice
the temperatures. The sharp lines on the left indicate bound states. For
comparison, experimental data are marked by arrows for bound peak
locations and by triangles for the continuum.  \cite{signconvention}.}
\label{fig:4}
\end{figure}

Figure~4 addresses recent data \cite{Ketterle4} associated with 1-3
pairing and RF excitation from state 3 to state 2.  The calculations of
$I(\nu)$ shown in the (black) solid curves in all six panels were
performed with experimental parameters, and should be compared with
Fig.~4 of Ref.~\cite{Ketterle4}. To help in the comparison a number of
data points (normalized to the same peak height) have been inserted.
The sharp bound states will, in the data, be broadened both
instrumentally and from limited spatial and energy resolution. Except
for a slight broadening which we have ignored here, our calculated black
solid curves, which incorporate final state effects, can be seen to be
in semi-quantitative agreement with experiment.  We anticipate that at
higher $T$ (red dashed lines), the negative $\nu$ continuum states
should start to become apparent.  Despite the presently good agreement,
we feel the ultimate test of any theory must involve a test of its
predictions, such as those shown here.

At unitarity, the best way to measure $\Delta(T)$ is using the sum rule
in Eq.~(\ref{eq:14}) and its experimental counterpart.  Together with
the $\nu > 0 $ continuum threshold which appears at
$\sqrt{\Delta^2+\mu^2}-\mu$, one can also determine $\mu$ and hence the
factor $\beta$.  This analysis is possible only in the presence of final
state effects.  Because $I(\nu)$ at general $T$ depends on the total
pairing gap $\Delta(T)$, the size of the order parameter and pseudogap
cannot be separately inferred (except when the order parameter vanishes
above $T_c$).

This work is supported by Grants NSF PHY-0555325 and NSF-MRSEC
DMR-0213745. We thank S. Basu, C. Chin, S. Jochim, and E. Mueller for
helpful discussions.

\vspace*{-1.5ex}

\bibliographystyle{apsrev}

\begin{thebibliography}{20}
\bibitem[{mea()}]{mean-field-footnote}
\bibinfo{note}{Here we extend this approximation above the actual $T_c$ as
  well, to estimate $\Delta$ and $\mu$.}
\bibitem[{sig()}]{signconvention}
\bibinfo{note}{Note that Ref.~\cite{Ketterle4} has a different sign convention
  for $\nu$.}
\end{thebibliography}

\end{document}